\documentclass[twocolumn,nofootinbib,amsmath,amssymb,aps,prd]{revtex4-1}

\usepackage{graphicx}
\usepackage[caption=false]{subfig}

\usepackage{amsmath,amssymb}
\usepackage{amsfonts,amssymb,mathrsfs}

\usepackage[bookmarks=false,pdfstartview=FitH]{hyperref}
\usepackage[all]{hypcap}

\def\be{\begin{equation}}
\def\ee{\end{equation}}
\def\nn{\nonumber}
\def\f{\frac}
\def\tf{\tfrac}
\def\sgn{{\rm sgn}}
\def\pl{{\rm Pl}}
\def\lp{\ell_\pl}
\def\b{\bar}
\def\d{\dot}
\def\h{\hat}
\def\t{\tilde}
\def\v{\vec}
\def\wh{\widehat}

\def\dd{{\rm d}}

\def\de{\delta}
\def\ep{\epsilon}
\def\ga{\gamma}
\def\la{\lambda}
\def\om{\omega}
\def\ve{\varepsilon}

\def\vt{\vartheta}
\def\De{\Delta}

\def\mH{\mathcal{H}}
\def\mO{\mathcal{O}}
\def\mP{\mathcal{P}}
\def\mC{\mathcal{C}}
\def\mN{\mathcal{N}}
\def\mV{\mathcal{V}}
\def\oe{\mathring{e}}
\def\ow{\mathring{\omega}}
\def\oq{\mathring{q}}

\def\omu{\mathring{\mu}}
\def\oepsilon{\mathring{\epsilon}}
\def\oxi{\mathring{\xi}}
\def\lo{\ell_o}

\def\bra{\langle}
\def\ket{\rangle}
\def\tr{{\rm Tr}}

\usepackage{color}

\begin{document}

\pagestyle{plain}

\title{Anisotropic loop quantum cosmology with self-dual variables}

\author{Edward Wilson-Ewing} \email{wilson-ewing@aei.mpg.de}
\affiliation{Max Planck Institute for Gravitational Physics (Albert Einstein Institute),\\
Am M\"uhlenberg 1, 14476 Golm, Germany, EU}

\begin{abstract}

A loop quantization of the diagonal class A Bianchi models starting from the complex-valued self-dual connection variables is presented in this paper.  The basic operators in the quantum theory correspond to areas and generalized holonomies of the Ashtekar connection and the reality conditions are implemented via the choice of the inner product on the kinematical Hilbert space.  The action of the Hamiltonian constraint operator is given explicitly for the case when the matter content is a massless scalar field (in which case the scalar field can be used as a relational clock), and it is shown that the big-bang and big-crunch singularities are resolved in the sense that singular and non-singular states decouple under the action of the Hamiltonian constraint operator.

\end{abstract}

\maketitle

\section{Introduction}
\label{s.intro}

The introduction of the self-dual connection variables in general relativity \cite{Ashtekar:1986yd, Ashtekar:1987gu} raised the possibility of developing a non-perturbative theory of quantum gravity based on the quantization techniques of gauge theories \cite{Jacobson:1987qk, Rovelli:1989za}.  However, there are two major obstacles that arise in the attempt to perform the canonical quantization of general relativity based on the complex-valued $SL(2, \mathbb{C})$ Ashtekar connection: (i) the fundamental operators of the theory must satisfy non-trivial reality conditions, and (ii) the measure for generalized $SL(2, \mathbb{C})$ connections is non-compact, with no known regulator available.

A major step forward was therefore made possible when it was shown that the real-valued $SU(2)$ Ashtekar-Barbero connection could be used instead \cite{Barbero:1994ap}, and indeed the use of these variables has led to significant progress in loop quantum gravity, most particularly results showing that there is a unique cyclic representation of the kinematical Hilbert space of LQG that is invariant under spatial diffeomorphisms \cite{Lewandowski:2005jk, Fleischhack:2004jc} (although see also \cite{Koslowski:2011vn, Dittrich:2014wpa}) as well as candidate definitions of the Hamiltonian constraint operator \cite{Thiemann:1996aw, Thiemann:2003zv, Domagala:2010bm, Husain:2011tk, Alesci:2015wla}.

Nonetheless, despite these important results, there are some drawbacks associated to the Ashtekar-Barbero variables.  First, the Ashtekar-Barbero variables require the introduction of the real-valued Immirzi parameter $\ga$ \cite{Immirzi:1996di}, a parameter that has no classical analog in standard general relativity, but appears in the spectra of operators corresponding to geometric observables \cite{Ashtekar:1996eg, Ashtekar:1997fb, Bianchi:2010gc, Thiemann:1996at, Bianchi:2008es, Ma:2010fy}.  (Note that the self-dual variables are recovered for $\ga = \pm i$.)  Second, the Hamiltonian constraint becomes more complicated, in that it contains an additional term whose presence necessarily leads to additional quantization ambiguities.  Third, the Ashtekar-Barbero connection is not a true space-time connection: while it transforms as a connection under spatial diffeomorphisms, it does not under time-like diffeomorphisms \cite{Samuel:2000ue, Alexandrov:2001wt}.

More recently, it has been shown that a black hole entropy of $S = A / 4 G \hbar$ is obtained after performing an analytic continuation of $\ga \to i$ \cite{Frodden:2012dq, Bodendorfer:2012qy, Han:2014xna, Achour:2014eqa}.  This result may indicate that an Immirzi parameter of $\ga=i$ (which is to say, self-dual variables) captures the correct black hole physics, and that it may be important to set the Immirzi parameter to $\ga=i$ in full LQG as well.  There are also a number of interesting studies on various other aspects of black hole physics in LQG which lead to similar conclusions \cite{Bodendorfer:2013hla, Pranzetti:2013lma, Geiller:2014eza, Carlip:2014bfa}.

These recent results in black hole physics, together with the drawbacks of the Ashtekar-Barbero connection described above, suggest that it may be fruitful to reexamine the two main obstacles to a well defined canonical theory of quantum gravity based on the self-dual Ashtekar variables, namely the reality conditions and the measure problem.  A first step in this direction is to study the canonical quantization of symmetry-reduced models where the reality conditions become simpler, and where the measure problem is typically avoided.  To date, the canonical quantization in terms of self-dual variables has been achieved for two families of symmetry-reduced space-times: the spherically symmetric and asymptotically flat vacuum Schwarzschild space-times \cite{Thiemann:1992jj}, and the homogeneous and isotropic Friedmann-Lema\^itre-Robertson-Walker (FLRW) cosmological space-times with a massless scalar field \cite{Wilson-Ewing:2015lia}.  (As an aside, note that an alternative approach to studying quantum cosmology with $\ga = i$ is to quantize in terms of the Ashtekar-Barbero variables and then analytically continue $\ga \to i$ \cite{Achour:2014rja}.)  In this paper, I will consider the canonical quantization in terms of self-dual variables of diagonal type A Bianchi space-times, which allow for the presence of anisotropies.  The restriction to type A Bianchi models (which will be defined below in Sec.~\ref{ss.def-b}) is necessary as these are the only ones for which a Hamiltonian formulation is known, and I only consider diagonal models for the sake of simplicity.  Note that this is a large family of space-times: diagonal type A Bianchi models include Bianchi space-times of type I, type II, type VIII and type IX.

While this work will extend in as straightforward a manner as is possible a number of results obtained for the FLRW space-times in \cite{Wilson-Ewing:2015lia}, it shall become clear that the presence of anisotropies significantly complicates the form of the reality conditions.  For the spatially flat and closed FLRW space-times (the ones studied in \cite{Wilson-Ewing:2015lia}), the reality conditions take an especially simple form since the spin-connection does not depend on the densitized triads.  This is no longer the case for the Bianchi space-times: now the spin-connection will depend in a non-trivial fashion on the densitized triads, and this will make the implementation of the reality conditions more difficult.

Furthermore, beyond the goal of exploring how the reality conditions can be imposed in a more general context, studying the quantum cosmology of the Bianchi models is also important for an additional reason: the Belinskii-Khalatnikov-Lifshitz (BKL) conjecture claims that, as a generic space-like singularity is approached, neighbouring points decouple in the sense that space-like derivatives are negligible compared to time-like derivatives \cite{Belinsky:1970ew, Belinsky:1982pk, Berger:2002st}.  Then, the dynamics at each point are well approximated by the ordinary differential equations that govern the evolution of the homogeneous Bianchi space-times.  Thus, understanding the dynamics of the Bianchi space-times in regions near where the space-time would become singular in classical general relativity (i.e., where quantum gravity effects are expected to become important) may also give important insights into the dynamics of generic space-times in regions where general relativity would predict a space-like singularity to arise.

Due to the importance of the BKL conjecture, the Bianchi space-times have already been studied in some detail in a variety of approaches, in particular in the loop quantum cosmology based on the Ashtekar-Barbero connection variables (which I shall call `standard LQC' in this paper).  The results of standard LQC for the Bianchi space-times can be found in the reviews \cite{Bojowald:2008zzb, Ashtekar:2011ni, Banerjee:2011qu} and the many references therein; since the object of this paper is a loop quantization for the Bianchi space-times using self-dual variables, many of the techniques developed in, and the results obtained from, studies of the standard LQC of the Bianchi models will be very useful here.

The outline of the paper is as follows: there is a brief review of the Bianchi space-times in general relativity and in particular a description of their Hamiltonian framework in terms of the self-dual Ashtekar variables in Sec.~\ref{s.bianchi}.  Then the canonical quantization is performed in Sec.~\ref{s.qc}: the reality conditions are imposed via an appropriate choice of the inner product for the kinematical Hilbert space in Sec.~\ref{ss.kin}, the Hamiltonian constraint operator is defined in Sec.~\ref{ss.ham}, and the resulting self-dual LQC is compared to standard LQC in Sec.~\ref{ss.comp}.  Then, the effective equations are briefly presented in Sec.~\ref{s.eff} and there is a discussion in Sec.~\ref{s.disc}.

\section{Bianchi Space-times}
\label{s.bianchi}

The Bianchi cosmologies are four-dimensional space-times that are spatially homogeneous and whose symmetry group is simply transitive.  The Bianchi models can be classified in terms of their symmetry group, and this classification is reviewed in Sec.~\ref{ss.def-b}.  In Sec.~\ref{ss.sym}, a fiducial cell is introduced in order to regulate integrals for the case of non-compact spaces, and the discrete parity symmetries of the Bianchi space-times are also described.  Finally, the Hamiltonian framework for the diagonal class A Bianchi models in terms of the self-dual variables is reviewed in Sec.~\ref{ss.self-dual}.  This section only covers the material that will be directly necessary for the remainder of the paper; significantly more information concerning the Bianchi cosmologies is given in, e.g., \cite{Ryan-Shepley, Jantzen:2001me}.

\subsection{Classification of the Bianchi Models}
\label{ss.def-b}

The Bianchi space-times are spatially homogeneous, with a three-dimensional isometry group spanning the spatial surface $\Sigma$ that can be parametrized by three linearly independent Killing vectors $\oxi^a_j$.  A basis of fiducial triads $\oe^a_j$ can be generated by requiring that their Lie derivative with respect to the Killing vector fields vanishes, $[\oe_j, \oxi_k]^a = 0$.

These fiducial triads satisfy the relations
\be
[\oe_k, \oe_l]^a = C^j{}_{kl} \oe^a_j,
\ee
while the fiducial co-triads $\ow_a^j$ ---that are dual to the fiducial triads--- satisfy
\be \label{cjkl}
\dd \ow^j + \f{1}{2} C^j{}_{kl} \, \ow^k \wedge \ow^l = 0,
\ee
where the $C^j{}_{kl} = C^j{}_{[kl]}$ are the structure constants of the Bianchi model.

The Bianchi models can be separated into two groups: class A where $C^j{}_{jk} = 0$, and class B where the trace of the structure constants does not vanish.  For Bianchi models of class A, it is possible to choose the fiducial triads in such a way so that the structure constants are entirely determined by three constants $n^j$ \cite{Jantzen:2001me}:
\be \label{def-n}
C^j{}_{kl} = n^j \oepsilon^j{}_{kl}, \quad {\rm no~sum~over~}j,
\ee
with $n^j = 0, \pm1$.  Here $\oepsilon_{jkl}$ is totally antisymmetric with $\oepsilon_{123} = 1$.  Table~\ref{table-n} shows the value of the $n^j$ for each class A Bianchi model.

\begin{table} \label{table-n}
\begin{tabular*}{0.4\textwidth}{@{\extracolsep{\fill}} |l|r@{}l|r r|r r|} \hline
Bianchi Model $\qquad$ & $n^1\!\!\!$ & & $n^2\!\!\!$ & & $n^3\!\!\!$ & \\ \hline
Type I & 0 & & 0 & & 0 & \\ \hline
Type II & 1 & & 0 & & 0 & \\ \hline
Type VI$_0$ & 1 & & -1 & & 0 & \\ \hline
Type VII$_0$ & 1 & & 1 & & 0 & \\ \hline
Type VIII & 1 & & 1 & & -1 & \\ \hline
Type IX & 1 & & 1 & & 1 & \\ \hline
\end{tabular*}
\caption[]{\footnotesize \hangindent=10pt
The values of $n^j$ for class A Bianchi models \cite{Jantzen:2001me}.}
\end{table}

The standard Hamiltonian treatment fails for class B Bianchi models \cite{MacCallum:1972er}, and there is no known canonical framework for these space-times.  For this reason only Bianchi models that are of class A will be considered in this paper.  Furthermore, for the sake of simplicity only diagonal models will be considered, in which case the space-time metric has the form
\be \label{diag}
ds^2 = -N(t)^2 dt^2 + \sum_j a_j(t)^2 (\ow^j)^2,
\ee
where $N(t)$ is the lapse function and $a_j(t)$ are the three directional scale factors.  Requiring that the metric have a diagonal form can always be achieved in the Bianchi models type VIII and IX via a careful choice for the $\ow_a^j$.  On the other hand, for the other class A Bianchi models imposing diagonality does correspond to a mild loss of generality in the sense that this restriction corresponds to fixing certain constants of motion \cite{Ashtekar:1992np}.

Given the line element \eqref{diag}, for diagonal class A Bianchi models the physical triads and co-triads are related to the fiducial ones via
\be \label{def-omega}
e^a_j = \f{1}{a_j(t)} \oe^a_j, \quad
\om_a^j = a_j(t) \ow_a^j, \quad {\rm no~sum~over~}j,
\ee
which define the spatial metric (at some fixed time) by
\be
q_{ab} = \om_a^j \om_b^j \, \de_{jk},
\ee
and therefore the problem of determining the dynamics of a diagonal class A Bianchi space-time has been reduced to solving for the time evolution of the three directional scale factors $a_j(t)$.

\subsection{Integrals and Parity Symmetries}
\label{ss.sym}

Before presenting the Hamiltonian framework for the diagonal class A Bianchi models, it is necessary to ensure that integrals are well defined (which is not automatic in non-compact homogeneous spaces), and also to understand how the basic variables change under a parity transformation so that this discrete symmetry can later be properly encoded in the quantum theory.  I will begin with the first point.

Depending on the Bianchi model, different topologies are allowed.  While topological and global aspects of the Bianchi space-times are not the main focus of this paper, it is important to differentiate between compact and non-compact spaces.  This distinction is necessary since the Hamiltonian that will be introduced in Sec.~\ref{ss.self-dual} is in fact the integral of a Hamiltonian density over the spatial surface, and integrals evaluated in homogeneous spaces (where there is no falloff at infinity) that are non-compact necessarily diverge.  On the other hand, if the space is compact, then the integral is finite.

Therefore, in compact spaces integrals can be evaluated over the entire space, while for non-compact spaces the integrals must be restricted to a finite region of the space.  This finite region of $\Sigma$ will be called the fiducial cell and denoted by $\mV$.  The fiducial cell acts as an infrared regulator, and once the quantum theory is defined, this regulator can be removed by taking the limit of $\mV \to \Sigma$.  (The classical theory is independent of the choice of $\mV$, but there do exist some subtle effects in the quantum theory concerning the amplitude of quantum fluctuations \cite{Rovelli:2013zaa}.  For this reason the regulator must be removed in the quantum theory in order to recover truly global observables that are not restricted to a finite subregion of $\Sigma$.  In the $\mV \to \Sigma$ limit, the quantum fluctuations of global observables will be negligible for sharply peaked states, but effects coming from the underlying Planck-scale quantum geometry will remain.  It appears necessary to go beyond the minisuperspace approximation in order to fully understand the effect of quantum fluctuations in non-compact homogeneous space-times.)

It is useful to define the quantities
\be \label{vo}
V_o = \int_\mV \sqrt{\oq}, \quad \lo = (V_o)^{1/3},
\ee
corresponding to the volume of the fiducial cell (or the spatial surface for compact spaces) with respect to the metric $\oq_{ab} = \ow_a^j \ow_b^k \de_{jk}$, and its cube root.  From now on, in order to simplify the notation, $\mV$ will represent the fiducial cell for non-compact spaces, and the entire spatial surface $\Sigma$ for compact spaces.  Due to homogeneity, the result of any integral over $\mV$ with respect to the fiducial metric is equivalent to multiplying the integrand by $V_o / \sqrt{\oq}$.

Thus, the introduction of the fiducial cell ensures that integrals are well defined, and acts as an infrared regulator in non-compact spaces.

The second topic of this part is parity transformations.  A parity transformation flips the orientation of one or several of the physical triads while leaving all fiducial quantities invariant.  While the metric does not depend on the orientation or the handedness of the triads (and so the physics is invariant under these parity transformations), the basic variables of the quantum theory will change under a parity transformation and therefore it is important to understand them and, ultimately, to include them as a discrete symmetry in the quantum theory.

Since the sign of the directional scale factors $a_j(t)$ encodes the orientation of the triads, the parity transformations act by changing the sign of the relevant directional scale factor while leaving all of the fiducial structures like $\oe^a_k$ and $\ow_a^j$ invariant.  For example, the parity transformation $\Pi_1$ that sends $e^a_1 \to -e^a_1$ acts on the $a_j$ as
\be \label{cl-parity}
\Pi_1 (a_1) = -a_1, \quad \Pi_1 (a_2) = a_2, \quad \Pi_1 (a_3) = a_3.
\ee
As already mentioned, the metric \eqref{diag} is clearly left invariant under parity transformations.  It is useful to define
\be
\ve_j = \sgn(a_j),
\ee
and then the $\ve_j$ can be used as a shorthand to denote the orientation of the triads.

An important point is that the spatial volume 3-form $\epsilon_{abc}$ is invariant under such a parity transformation: the integral of some function over $\mV$ does not depend on the choice of the triads (and therefore does not depend on their orientation either).  This is due to the fact that the orientation of the manifold and the orientation of the triads $e^a_i$ are not the same thing and may be opposite.  Here, the orientation of the manifold is fixed, while the orientation of the triads is what is reversed under the action of the parity operators $\Pi_j$.  Hence, the spatial volume 3-form $\epsilon_{abc}$ is fixed.

However, since the 3-form in internal indices $\epsilon_{jkl}$ (not to be confused with the antisymmetric tensor $\oepsilon_{jkl}$ appearing in the structure constants which is invariant under all transformations) is related to the volume 3-form as
\be
\epsilon_{jkl} = \epsilon_{abc} e^a_j e^b_k e^c_l,
\ee
it follows that under a parity transformation $\epsilon_{jkl}$ does change sign,
\be
\Pi_j(\epsilon_{klm}) = -\epsilon_{klm}.
\ee
This suggests the definition
\be
\ve = \epsilon_{123},
\ee
where $\ve = \pm1$, and so $\Pi_j(\ve) = -\ve$.  Given the transformation properties of $\ve$ under the $\Pi_j$ and using the condition that the orientation of $\epsilon_{abc}$ and $\epsilon_{jkl}$ should agree for right-handed triads (i.e., when all $\ve_j$ are positive and the orientation of the triads agrees with that of the manifold), it follows that
\be
\ve = \ve_1 \ve_2 \ve_3.
\ee
Thus, under a parity transformation $\Pi_j$, the orientation of that triad is flipped, which corresponds to changing the sign of $\ve_j$.  From this discussion, it follows that the Ashtekar connection, the densitized triad and the spin-connection (which will all be introduced in the next subsection) all change sign under a parity transformation.  It can be checked that the reality conditions transform properly under the action of $\Pi_j$, and it will also be necessary to appropriately incorporate the effect of the parity transformations in the quantum theory.

\subsection{Self-Dual Variables}
\label{ss.self-dual}

The complex-valued self-dual variables are the self-dual Ashtekar connection,
\be
A_a^j = \Gamma_a^j + i K_a^j,
\ee
which is constructed from the spin-connection $\Gamma_a^j$ and the extrinsic curvature $K_a^j = K_{ab} e^b_j$, and the conjugate variable to $A_a^j$, the densitized triad
\be
E^a_j = \sqrt{q} \, e^a_j,
\ee
which is composed of the determinant $q$ of the spatial metric and the physical triads $e^a_j$ defined in \eqref{def-omega}.  Since the Ashtekar connection is complex-valued, in order to recover (real-valued) general relativity, it is necessary to impose the two reality conditions
\be
A_a^j + \left( A_a^j \right)^\star = 2 \, \Gamma_a^j, \quad E^a_j E^b_k \de^{jk} > 0.
\ee
These are the basic variables that the remainder of the paper will be based upon.  For a more detailed introduction to self-dual variables, see, e.g., \cite{Ashtekar}.

For diagonal class A Bianchi models, the Ashtekar connection and the densitized triads can be parametrized as \cite{Bojowald:2003md}
\be \label{def-cp}
A_a^j = \f{c_j}{\lo} \ow_a^j, \quad E^a_j = p_j \f{\sqrt{\oq}}{\lo^2} \oe^a_j,
\quad {\rm no~sum~over~}j,
\ee
and then, for these variables, the symplectic structure of self-dual general relativity induces the Poisson brackets
\be
\{c_j, p_k\} = i \cdot 8 \pi G \, \delta_{jk}.
\ee

It is helpful to relate the $c_j$ and $p_j$ to the more familiar geometrodynamical variables in terms of $a_j(t)$ and $\d a_j(t)$, where the dot denotes differentiation with respect to proper time.  From \eqref{def-omega} and $\sqrt{q} = |a_1 a_2 a_3| \sqrt{\oq}$, it immediately follows that the $p_j$ are related to the directional scale factors as, e.g., $p_1 = \ve_1 |a_2 a_3| \lo^2$ and cyclic permutations thereof.  Clearly, the sign of $p_j$ determines the orientation of $e^a_j$, and changes under the parity transformation $\Pi_j$.

Relating $c_j$ to the geometrodynamical variables requires two steps.  First, the spin-connection components are, e.g., \cite{Bojowald:2003md, Bojowald:2003xf}
\begin{align}
\Gamma_a^1 &= -\epsilon^{1jk} e^b_j \left( \partial_{[a} \om_{b]k} + \f{1}{2}
e^c_k \om_a^l \partial_{[c} \om_{b]l} \right) \nn \\ \label{spin-con}
&= \f{\ve}{2} \left( n^2 \, \f{p_3}{p_2} + n^3 \, \f{p_2}{p_3} - n^1 \, \f{p_2 p_3}{p_1^2} \right) \ow_a^1 =: \Gamma^1 \ow_a^1,
\end{align}
where the shorthand $\Gamma_a^j = \Gamma^j \ow_a^j$ (no sum over $j$) has been introduced.  The other components of the spin-connection can be obtained via cyclic permutations.  Importantly, the spin-connection components are diagonal with respect to the fiducial co-triads $\ow_a^j$: otherwise the parametrization of the Ashtekar connection in \eqref{def-cp} would fail.  This is a property of class A Bianchi models that greatly simplifies the Hamiltonian treatment, as well as the quantization procedure.

Second, it is easy to check that in general relativity the extrinsic curvature is given by $K_a^j = \d a^j \ow_a^j$, where there is no sum over $j$ and the dot denotes a derivative with respect to the proper time $t$.  (Note however that the relation between the extrinsic curvature and the proper time derivatives of the directional scale factors is more complicated in LQC, even in the effective theory.)  The combination of these two results shows how each of the $c_j$ can be related to the directional scale factors and their time derivatives in classical general relativity as $c_j = \lo(\Gamma_j + i \cdot \d a_j)$.

In order to express the spin-connection in a simpler way, it is helpful to introduce
\be
r_1 = \left| \f{p_2 p_3}{p_1} \right|, \quad
r_2 = \left| \f{p_1 p_3}{p_2} \right|, \quad
r_3 = \left| \f{p_1 p_2}{p_3} \right|,
\ee
in order to define
\be \label{def-s}
s = n^j r_j.
\ee
Then, the spin-connection is given by
\be
\Gamma^j = \f{1}{2} \cdot \f{\partial s}{\partial p_j} = - \f{i}{16 \pi G} \{c_j, s\}.
\ee
Writing the spin-connection in this way not only leads to a relatively simple form of the reality conditions in the classical theory,
\be \label{real}
c_j + c_j^\star = - \f{i \lo}{8 \pi G} \, \{c_j, s\}, \qquad
p_j^\star = p_j,
\ee
but will also give some important insight into how the reality conditions can be imposed in the quantum theory.

The next step is to determine the dynamics of the $c_j$ and $p_j$, which requires the construction of the Hamiltonian constraint which is composed of the scalar, diffeomorphism and Gauss constraints.  It is easy to check that when the self-dual variables have the form \eqref{def-cp}, the Gauss and diffeomorphism constraints are automatically satisfied, and only the scalar constraint is left,
\be
\mH = \left[ \f{E^a_j E^b_k}{16 \pi G \sqrt{q}} \epsilon^{jk}{}_l F_{ab}{}^l + \f{\pi_\phi^2}{2 \, \sqrt{q}} \right] \approx 0,
\ee
where the `$\approx 0$' indicates that the constraint must vanish for physical solutions.  Here the matter content has been chosen to be a massless scalar field $\phi$, and its conjugate momentum is denoted by $\pi_\phi$.

Since the Gauss and diffeomorphism constraints vanish, the Hamiltonian constraint is simply $\mC_H = \int N \mH$, with $N$ being the lapse.  In terms of the $c_j$ and $p_j$, $\mC_H$ is
\begin{align}
\mC_H &= \f{N}{8 \pi G \sqrt{|p_1 p_2 p_3|}} \Bigg[ p_1 p_2 \left( c_1 c_2 - n^3 \lo \ve c_3 \right) \nn \\
& \quad\: + p_2 p_3 \left( c_2 c_3 - n^1 \lo \ve c_1 \right)
+ p_1 p_3 \left( c_1 c_3 - n^2 \lo \ve c_2 \right) \Bigg] \nn \\ \label{cl-ham}
& \quad\: + \f{N p_\phi^2}{2 \sqrt{|p_1 p_2 p_3|}} \approx 0,
\end{align}
where $\pi_\phi = \sqrt{|q|} \, \d \phi = p_\phi \sqrt{\oq} / V_o$.  With this definition for $p_\phi$, the Poisson bracket for the massless scalar field is given by $\{\phi, p_\phi\} = 1$.  Note that all terms $\ve^2$ have been set to 1.

The Hamiltonian constraint generates the dynamics of the Bianchi models; for any observable $\mO$,
\be
\f{d \mO}{dT} = \{\mO, \mC_H\},
\ee
where $T$ is the time variable related to proper time via $N dT = dt$.  For example, taking $N=1$,
\be
\d p_1 = \f{-i p_1}{\sqrt{|p_1 p_2 p_3|}} \left[ p_2 c_2 + p_3 c_3 - \lo n^1 \ve \, \f{p_2 p_3}{p_1} \right],
\ee
(recall that the dot denotes $d/dt$).  Clearly, the time derivatives of the other $p_j$ can be obtained via cyclic permutations, and time derivatives of other observables of interest can be obtained by the same procedure.

Finally, generalized holonomies play an important role in the quantum theory.  In the previous treatment of the self-dual LQC of FLRW space-times \cite{Wilson-Ewing:2015lia}, a family of `generalized holonomies' parametrized by $\alpha \in \mathbb{C}$ were introduced,
\be \label{def-gh}
h = \mP \exp \left[ \int_{\rm edge} \!\!\!\!\!\! \alpha A_a \right],
\ee
with $A_a := A_a^j \, \sigma_j / 2i$, where the $\sigma_j$ are the Pauli matrices; standard holonomies are recovered for $\alpha = 1$.  This extension was necessary since it was impossible to define operators corresponding to standard holonomies in the kinematical Hilbert space of self-dual LQC for FLRW space-times.  Rather, only generalized holonomies with $\alpha$ completely imaginary were well defined.  This will turn out to be the case for Bianchi models as well, as shall be shown in Sec.~\ref{ss.kin}.

At this point it is worth pointing out two properties that generalized holonomies have in common with standard holonomies.  First, generalized holonomies are elements of $SL(2, \mathbb{C})$.  The definition \eqref{def-gh} has a simple form due to the representation of the self-dual $\mathfrak{sl(2,\mathbb{C})}$ connection $A_a^j$ being a three-dimensional complex vector space, and thus it is clear that if $A_a \in \mathfrak{sl(2,\mathbb{C})}$, so is $\alpha \, A_a$ for $\alpha \in \mathbb{C}$.  (Note that defining generalized holonomies for other Lie algebras may not be as straightforward.)  Second, the composition and inversion rules for generalized holonomies are the same as for standard holonomies.  These properties hold generally, and not just for homogeneous space-times.  Note however that generalized holonomies do not transform in the same way as standard holonomies under gauge transformations.

Of course, the actual form of the generalized holonomies will simplify in Bianchi space-times.  Due to their spatial homogeneity, it is sufficient to calculate the generalized holonomies that are tangent to the fiducial triads.  A generalized holonomy of the Ashtekar connection tangent to $\oe^a_j$ and of length $\omu \lo$ with respect to the fiducial metric is
\begin{align}
h_j(\omu) &= \exp \left( \int_0^{\omu \lo} \!\! A_a^j \oe^a_j \, \alpha \sigma_j /2i \right) \nn \\ &
%\quad {\rm no~sum~over~}j \nn \\ &
%= \exp \left( \alpha \omu c_j \sigma_j / 2i\right) \nn \\ &
\label{hol}
= \cosh \left( \f{\alpha \omu c_j}{2i} \right) \mathbb{I} + \sinh \left( \f{\alpha \omu c_j}{2i} \right) \sigma_j,
%\quad {\rm no~sum~over~}j
\end{align}
where there is no sum over $j$ in either of the lines.  Since the only dependence on $c_j$ in these expressions is in the hyperbolic trigonometric functions, it will be sufficient to define operators corresponding to exponentials of $\alpha \omu c_j / 2i$ in order to represent generalized holonomies in the quantum theory.

An important relation for the quantum theory is
\be \label{hol->c}
c_j = \lim_{\omu \to 0} i \f{\tr [h_j(\omu) \sigma_j]}{\alpha \omu}, \qquad {\rm no~sum~over~} j,
\ee
which shows that the phase space variable $c_j$ can easily recovered from the holonomies $h_j(\omu)$.  This will allow for the definition of a non-local connection operator in terms of a generalized holonomy $h_j(\omu)$ of an appropriate (Planck-sized) length, as shall be explained in Sec.~\ref{ss.ham}.

\section{The Quantum Theory}
\label{s.qc}

Using the function $s(p_j)$ determined by the Bianchi model structure constants in \eqref{def-s}, it is possible to proceed with the canonical loop quantization of all diagonal class A Bianchi models in one go.  This section is split in three parts: in the first, the kinematical Hilbert space is defined and the reality conditions of the fundamental operators of the theory are imposed through the choice of the inner product; in the second, the Hamiltonian constraint operator is constructed and some of its properties are studied; and in the third, the resulting theory of self-dual LQC is compared to standard LQC.

\subsection{The Kinematical Hilbert Space}
\label{ss.kin}

The kinematical Hilbert space $H$ is given by the tensor product of the kinematical Hilbert space of the gravitational sector $H_g$ and of the matter sector $H_m$,
\be
H = H_g \otimes H_m.
\ee

For $H_g$, the second reality condition in \eqref{real} suggests that it may be convenient to use $|p_1, p_2, p_3\ket$ as a basis for the gravitational sector of the kinematical Hilbert space, with $p_j \in \mathbb{R}$ due to the reality condition.  For now, it shall be assumed that any single basis vector of this type is normalizable; this shall be shown to be the case below.

By definition, the $\h p_j$ operators act by multiplication on this basis, for example
\be \label{p-hat}
\h p_1 | \v p \, \ket = p_1 | \v p \, \ket,
\ee
where $|\v p \, \ket$ is shorthand for $|p_1, p_2, p_3 \ket$.  The other family of fundamental operators in the gravitational sector corresponds to generalized holonomies of $A_a^j$ along paths tangential to $\oe^a_j$ of length $\omu$, which following \eqref{hol} can be entirely expressed in terms of the shift operators
\be \label{shift}
\wh{e^{\mu c_1}} |p_1, p_2, p_3 \ket = |p_1 + 8 \pi G \hbar \mu, p_2, p_3 \ket,
\ee
where $\mu = \alpha \omu / 2i$.  Clearly, since the states $| \v p \, \ket \in H_g$ only for $p_j \in \mathbb{R}$, it follows that $\mu \in \mathbb{R}$ is a necessary condition for this shift operator to be well defined on $H_g$.  Since $\omu$ is real-valued by definition, it follows that $\alpha$ must be purely imaginary, and this shows that it is the same family of generalized holonomies that is well defined for the self-dual LQC of the FLRW space-times and of the Bianchi models.  In neither case are the standard holonomy operators well defined in self-dual LQC, and only operators corresponding to generalized holonomies with a purely imaginary $\alpha$ exist in their respective kinematical Hilbert spaces.  This suggests that, when working with the self-dual connection variables, holonomies are not appropriate operators in the quantum theory: instead one should work with generalized holonomies with imaginary $\alpha$.

Finally, the last fundamental operator to be defined is the inverse triad operator.  The starting point of the definition of $H_g$ was assuming the states $|\v p\, \ket$ to be normalizable basis vectors (again, this will be shown to be the case below).  Then, since $|0,0,0\ket$ is an element of $H_g$, the operator $(\h p_j)^{-1}$ is not well defined on $H_g$ and for this reason it is necessary to define an alternative inverse triad operator.  This can be done by adapting one of the Thiemann identities \cite{Thiemann:1996aw} to the symmetry-reduced phase space of LQC, but there is considerable freedom in this choice (see e.g.~\cite{Singh:2013ava} for a discussion concerning the quantization ambiguities related to the choice of the inverse triad operator in standard LQC; the same ambiguities arise in self-dual LQC).  In this paper, I shall define the inverse triad operators to be of the form
\be \label{inv}
\wh{\f{1}{p_1}} | \v p \, \ket =
\begin{cases}
0 & {\rm for~} ~ p_1 = 0 \\
\f{1}{p_1} | \v p \, \ket &{\rm otherwise},
\end{cases}
\ee
(and analogous definitions for the inverse triad operators corresponding to $p_2$ and $p_3$) for two reasons.  First, for any other known choice of inverse triad operator, the operator depends on global quantities and, for non-compact spaces, on the choice of the fiducial cell.  While this is especially problematic for the case of non-compact spaces since any dependence of physical quantities on the fiducial is unphysical, this dependence in non-compact spaces vanishes in the limit of $\mV \to \Sigma$.  However, once this limit has been taken, the resulting form of the inverse triad operator is necessarily \eqref{inv}.  (Note that while other choices of inverse triad operators are possible for compact spaces, it is simplest to take the same inverse triad operator for compact and non-compact spaces.)  Second, this choice of inverse triad operator is particularly convenient since the product of $\h p_j$ with its `inverse' \eqref{inv} is the identity everywhere, except for the basis vectors where $p_j=0$ which are annihilated by the two operators.

Following the same philosophy, the $\h r_j$ operators are defined following \eqref{p-hat} and \eqref{inv}, for example
\be \label{def-r-op}
\h r_1 | \v p \, \ket =
\begin{cases}
0 & {\rm if~any~} ~ p_j = 0, \\
\left| \f{p_2 p_3}{p_1} \right| | \v p \, \ket &{\rm otherwise}.
\end{cases}
\ee

With these three types of operators defined, the next step is to define the inner product in such a way that the reality conditions \eqref{real} are properly implemented in the quantum theory.  The reality conditions, in terms of the operators defined above, become
\be \label{real-p}
(\h p_j)^\dag = \h p_j.
\ee
and
\be \label{real-1}
\left(\wh{e^{\mu c_j}}\right)^\dag \sim e^{-\mu( c_j + \lo [c_j, s] / 8 \pi G \hbar )}.
\ee
Note that the Poisson bracket in \eqref{real} has been replaced by a commutator via $\{\cdot, \cdot\} \to [\cdot, \cdot]/i\hbar$, and recall also from the discussion above that $\mu \in \mathbb{R}$ and therefore $\b \mu = \mu$.  The $\sim$ in \eqref{real-1} indicates that the operator on the right-hand-side remains to be precisely defined; among other ambiguities it is necessary to choose a particular factor-ordering.  However, no matter the factor-ordering, due to the non-commutativity of the basic operators it will only be possible to impose the reality conditions up to leading order with quantum corrections of the order of $\hbar$ unavoidable.

The qualitative form of the second reality condition and the Baker-Campbell-Hausdorff equation (to leading order) suggest that the operator equation corresponding to the reality condition \eqref{real} could be written as
\be \label{real-2}
\left(\wh{e^{\mu c_j}}\right)^\dag = \wh{e^{-s \lo / 8 \pi G \hbar}} \; \wh{e^{-\mu c_j}} \; \wh{e^{s \lo / 8 \pi G \hbar}}.
\ee
While in principle it would be possible to include the higher order terms in the Baker-Campbell-Hausdorff equation, there is no need for this since there already exist operator-ordering ambiguities in the definition of this operator equation in any case, and the reality condition as written in \eqref{real-2} is already sufficient in order to recover the classical relation \eqref{real} for states that have support on $p_j$ much greater than the $\lp^2$.

Further support for this choice is given by the form this reality condition has in the quantum theory for the closed FLRW space-time \cite{Wilson-Ewing:2015lia},
\be
\left(\wh{e^{\mu c}}\right)^\dag = \wh{e^{-3 |p| \lo / 8 \pi G \hbar}} \; \wh{e^{-\mu c}} \; \wh{e^{3 |p| \lo / 8 \pi G \hbar}}. \nn
\ee
The closed FLRW space-time is obtained by starting from the Bianchi IX model and imposing isotropy.  In that case, the structure constants are $n^1 = n^2 = n^3 = 1$ and all of the directional variables are taken to be equal, $p = p_j, c = c_j$.  It is clear that for the closed FLRW cosmology, $s = 3|p|$.  Then, it is easy to see that replacing $3|p|$ by $s$ in the reality condition for FLRW space-times also gives precisely \eqref{real-2}.

In order to determine the appropriate inner product, it is again helpful to consider the self-dual LQC of closed FLRW space-times, where the inner product is
\be
\bra k | p \, \ket = e^{3 \lo |p| / 8 \pi G \hbar} \, \de_{k, p}, \nn
\ee
which suggests that the inner product for the anisotropic space-times in self-dual LQC should be
\be \label{inner}
\bra \v k \, | \v p \, \ket = e^{s(\v p) \lo / 8 \pi G \hbar} \, \de_{k_1, p_1} \de_{k_2, p_2} \de_{k_3, p_3},
\ee
where $\h s \, | \v p \, \ket = s(\v p \,) \, | \v p \, \ket$, with $\h r_j$ being defined in \eqref{def-r-op}, and $\de$ denotes the Kronecker delta.  With this inner product, all kets $|\v p\ket$ with $p_j \in \mathbb{R}$ are normalizable, and this verifies the assumption made at the beginning of this section.

Given this inner product, any normalized state in the Hilbert space has the form
\be
\psi(p) = \sum_{\v p \, \in \, \mathbb{R}^3} \! C_{\v p} \, |\v p \, \ket,
\ee
with $\sum_{\v p} e^{s(\v p) \lo / 8 \pi G \hbar} |C_{\v p}|^2 = 1$.

It is a straightforward task to show that with the inner product \eqref{inner}, the reality conditions encoded in the operator equations \eqref{real-p} and \eqref{real-2} hold.  First, the reality condition \eqref{real-p} is clearly satisfied by imposing that $H_g$ is spanned by the basis kets $| \v p \, \ket$ with $p_j \in \mathbb{R}$.  Second, using the definition of the inner product, it follows that \eqref{real-2} is also satisfied:
\begin{align}
\bra \v k \, | \wh{e^{-s \lo / 8 \pi G \hbar}} \wh{e^{-\mu c_j}} \wh{e^{s \lo / 8 \pi G \hbar}} | \v p \, \ket &= e^{(\De s) \lo / 8 \pi G \hbar} \bra \v k \, | \v p_- \ket \nn \\ 
&= e^{s(\v p) \lo / 8 \pi G \hbar} \, \de^3_{\v k, \v p_-} \nn \\
&= e^{s(\v p) \lo / 8 \pi G \hbar} \, \de^3_{\v k_+, \v p} \nn \\ \label{real-3}
&= \bra \v k \, | \big( \wh{e^{\mu c_j}} \big)^\dag | \v p \, \ket, 
\end{align}
where $|\v p_- \ket = \wh{e^{-\mu c_j}} |\v p \, \ket, |\v k_+\ket = \wh{e^{\mu c_j}} |\v k \, \ket$, and $\De s = s(\v p) - s(\v p_-)$.

To show that \eqref{real-2} correctly captures the classical reality condition \eqref{real}, in the relation
\be
e^{(\De s) \lo / 8 \pi G \hbar} \bra \v k \, | \wh{e^{-\mu c_j}} | \v p \, \ket = \bra \v k \, | \left( \wh{e^{\mu c_j}} \right)^\dag | \v p \, \ket
\ee
coming from the first and fourth lines of \eqref{real-3}, the $\De s$ term can be Taylor-expanded for $p_j$ much greater than the Planck area.  For example, for the shift operator with internal index $j=1$, $|\v p_-\ket = |p_1 - 8 \pi G \hbar \mu, p_2, p_3 \ket$ and, assuming that $p_1 \gg 8 \pi G \hbar \mu$ and $p_2, p_3 > 0$,
\begin{align}
\De s &= n^1 \!\! \left( \f{p_2 p_3}{p_1} - \f{p_2 p_3}{p_1 - 8 \pi G \hbar \mu} \! \right)
+ 8 \pi G \hbar \mu \! \left( \! n^2 \f{p_3}{p_2} + n^3 \f{p_2}{p_3} \! \right) \nn \\
&\approx 8 \pi G \hbar \mu \left( -n^1 \f{p_2 p_3}{p_1^2} + n^2 \f{p_3}{p_2} + n^3 \f{p_2}{p_3} \right),
\end{align}
where terms of the order $8 \pi G \hbar \mu p_2 p_3/p_1^3$ and smaller have been dropped; and the result is clearly the spin-connection component $\Gamma^1$ given in \eqref{spin-con} multiplied by $16 \pi G \hbar \mu$.  (It is easy to check that the correct result is also obtained if any number of the $p_j$ are negative, so long as $|p_1| \gg 8 \pi G \hbar \mu$.)  The same calculation clearly holds for $j=2$ and $j=3$ and thus, up to terms of the order of $G \hbar / p_j$ which are negligible in the semi-classical limit,
\be
e^{2 \lo \mu \Gamma^j} \bra \v k \, | \wh{e^{-\mu c_j}} | \v p \, \ket \approx \bra \v k \, | \left( \wh{e^{\mu c_j}} \right)^\dag | \v p \, \ket.
\ee
This shows that the operator equation \eqref{real-2} does indeed provide an appropriate form of the reality condition \eqref{real} for the quantum theory.

The last operators that remain to be defined on $H_g$ are the parity operators corresponding to the discrete symmetries \eqref{cl-parity}, which act as, for example,
\be
\wh \Pi_1 |p_1, p_2, p_3\ket = |-p_1, p_2, p_3\ket.
\ee
It is also convenient to define the $\h \ve_j$ operators as a shorthand for the operators corresponding to $\sgn(p_j)$,
\be
\h \ve_j = \wh{\sgn(p_j)}.
\ee
Since the orientation of the triads does not affect the classical theory, the wave functions $\psi(\v p)$ are required to be invariant under parity transformations, i.e.,
\be \label{parity}
\wh \Pi_j \psi(\v p) = \psi(\v p),
\ee
for all $j$.  This completes the definition of the kinematical Hilbert space for the gravitational sector.

To recap, the kinematical Hilbert space $H_g$ is spanned by the basis vectors $|\v p \, \ket$ with $\v p \in \mathbb{R}^3$, and the fundamental operators are the $\h p_j$ which act by multiplication, the shift operators \eqref{shift} and the inverse triad operators \eqref{inv}.  Then, the classical reality conditions are translated into the operator equations \eqref{real-p} and \eqref{real-2}, and these operator equations hold for the inner product \eqref{inner}.  Furthermore, wave functions are required to be invariant under the parity transformations \eqref{parity}.

Finally, the kinematical Hilbert space $H_m$ corresponding to the scalar field sector is the standard space of square-integrable functions $\chi(\phi)$ with respect to the Lebesgue measure $d\phi$, and the fundamental operators are
\be
\h \phi \, \chi(\phi) = \phi \, \chi(\phi), \qquad \h \pi_\phi \, \chi(\phi) = - i \hbar \f{d \chi(\phi)}{d \phi}.
\ee

\subsection{The Hamiltonian Constraint Operator}
\label{ss.ham}

In order to define the operator corresponding to the Hamiltonian constraint \eqref{cl-ham} for some choice of the lapse function, it is necessary to define an operator corresponding to the phase space variables $c_j$: while operators corresponding to $\pi_\phi$ and the $p_j$ (as well as inverse powers of them) are defined in the previous section, no operator corresponding directly to $c_j$ was introduced.

This is no accident as there is in fact no operator that directly corresponds to $c_j$ in the quantum theory: it can readily be checked that the action of the shift operator \eqref{shift} is not continuous with respect to the parameter $\mu$ and therefore infinitesimal shifts are not well defined.  (This is because the kinematical inner product depends on Kronecker delta functions which are not continuous in their arguments.)  Instead, it is necessary to express $c_j$ in terms of small but finite shift operators.  This follows from the definition of the kinematical Hilbert space which was constructed following the techniques of loop quantum gravity, and is analogous to the fact that (generalized) holonomies of the connection are the fundamental operators of the theory, not the connection itself.

Furthermore, the requirement that the $c_j$ variables must be represented in the quantum theory by holonomies of finite length can be understood to capture the Planck-scale discreteness of loop quantum gravity: given the discrete spectrum of geometrical observables in loop quantum gravity, it does not make sense to calculate the parallel transport of the connection along a path that is shorter than the Planck length.  Therefore, it is appropriate to define the non-local operator corresponding to the $c_j$ phase space variables using holonomies along edges of a minimum physical length determined by quantum gravity%
\footnote{There are two reasons for defining a non-local connection operator rather than a non-local field strength operator for the self-dual LQC of Bianchi space-times: (i) for generalized holonomies with a purely imaginary $\alpha$, the relation between the field strength $F_{ab}{}^k$ and the matrix elements $F_{ab}{}^A{}_B$ is much more complicated than for the connection \cite{Wilson-Ewing:2015lia}, and (ii) even in standard LQC, for Bianchi models with non-vanishing spatial curvature the field strength is not an almost-periodic function of the connection and therefore it is not known how to represent it as an operator in the quantum theory \cite{Ashtekar:2009um, WilsonEwing:2010rh}.}.
This minimum length $\ell_m$ is expected to be of the order of the Planck length, although in order to obtain a specific value it would be necessary to derive (self-dual) LQC from (self-dual) LQG.  (In the standard LQC of the FLRW and the Bianchi I space-times, it is the field strength that is expressed in terms of holonomies, and it is assumed that the area of the loop encircled by the holonomy is the smallest non-vanishing eigenvalue of the area operator in LQG \cite{Ashtekar:2006wn, Ashtekar:2006es, Ashtekar:2009vc}.  This suggests that taking $\ell_m$ to be the square root of the minimal non-zero eigenvalue of the area operator in self-dual LQG \cite{Rovelli:1994ge} would be a reasonable choice, but since a specific choice is not necessary in any case here, I will leave $\ell_m$ free.)

Following this reasoning, the non-local connection operator $\h c_j$ is obtained via the relation \eqref{hol->c}, with the difference that, instead of taking the limit of the path length to vanish, it is set equal to the minimal physical length $\ell_m$.

It is important to note that this minimal length is of course measured with respect to the physical metric $q_{ab}$ and not with respect to the fiducial metric $\oq_{ab}$.  Since the holonomies $h_j(\omu_j)$ defined in \eqref{hol} have a length of $\omu_j \lo$ with respect to $\oq_{ab}$, their physical length with respect to $q_{ab}$ is $|a_j| \omu_j \lo$ (no sum over $j$).  Thus, requiring that the physical path-length of the holonomy be $\ell_m$ (and recall also the relations between the scale factors and the $p_j$, e.g., $p_1 = \sgn(a_1) |a_2 a_3| \lo^2$) gives
\be
\ell_m = |a_j| \omu_j \lo \quad \Rightarrow \quad \omu_j = \ell_m \cdot \sqrt{ \left| \f{p_k p_l}{p_j} \right|},
\ee
where there is no sum over $j$ and it is understood that in the second equation $k$ and $l$ are the two indices different from $j$; for example, $\omu_1 = \ell_m \sqrt{|p_2 p_3 / p_1|}$.  This procedure gives a value of $\omu$ which corresponds to the standard `improved dynamics' loop quantization procedure of the Bianchi models \cite{Ashtekar:2009vc, Ashtekar:2009um, WilsonEwing:2010rh}.

The resulting non-local $\h c_j$ operator (ignoring factor-ordering ambiguities for now and dropping the hats on operators in order to avoid unnecessary clutter in the notation) is
\be
\h c_j = \f{e^{\la_m \sqrt{|p_j/p_k p_l|} c_j} - e^{-\la_m \sqrt{|p_j/p_k p_l|} c_j}}{2 \la_m \sqrt{|p_j/p_k p_l|}},
\ee
where $\la_m = \alpha \ell_m / 2i$, and again there is no sum over any of the internal indices and it is understood that $k \neq l$ are the two indices different from $j$.

Note however that the operator
\be \label{shift-1}
\mathfrak{N}_j^\pm = e^{\pm \la_m \sqrt{|p_j/p_k p_l|} c_j}
\ee
(where as above there is no sum over $j$, and $k, l$ are understood to be different from $j$ and each other) is not one of the simple shift operators defined in \eqref{shift} since the exponent includes $p_j$ terms and therefore it is not immediately obvious how it acts on a given state.  The action of this operator can be understood by introducing the variables
\begin{align}
b_j &= \sqrt{|p_j|} c_j, &{\rm no~sum~over~} j, \\
\vt_j &= \sgn(p_j) \sqrt{|p_j|}, ~&{\rm no~sum~over~} j, \\
V &= \vt_1 \vt_2 \vt_3.
\end{align}
Since the Poisson bracket of the variables $(b_j, \vt_k)$ is given by $\{b_j, \vt_k\} = i \cdot 4 \pi G \de_{jk}$, the operator $\mathfrak{N}_j^\pm$ can be understood to act as a shift operator where the wave function is shifted in the $\vt_j$ argument by an amount which depends on $\vt_k$ and $\vt_l$.  For example,
\be
\mathfrak{N}_1^\pm |\vt_1, \vt_2, \vt_3 \ket = |\vt_1 \pm 4 \pi G \la_m \cdot |\vt_2 \vt_3|^{-1}, \vt_2, \vt_3\ket.
\ee
Again, this is analogous to the `improved dynamics' prescription for Bianchi models in standard LQC \cite{Ashtekar:2009vc, Ashtekar:2009um, WilsonEwing:2010rh}.  Here the basis $|\v\vt\ket$ is simply a relabeling of $|\v p \ket$,
\be
|\vt_1, \vt_2, \vt_3 \ket_{\v\vt} = |\sgn(\vt_1) \vt_1^2, \sgn(\vt_2) \vt_2^2, \sgn(\vt_3) \vt_3^2 \ket_{\v p},
\ee
where the subscript denotes the label of the basis states.

This completes the definition of the non-local connection operator $\h c_j$ (up to factor-ordering choices), and with this it is now possible to construct and study the Hamiltonian constraint operator.

In order to do this, it is necessary to choose a lapse and a specific factor-ordering for the Hamiltonian constraint operator and then determine its action.  The Hamiltonian constraint operator contains a number of terms, which can be grouped in the following fashion:
\be \label{qham}
\wh{\mC}_H = \wh{\mC}_{(12)} + \wh{\mC}_{(23)} + \wh{\mC}_{(13)} + \lo n^j \wh{\mC}^n_j - \f{\hbar^2}{2} \partial_\phi^2.
\ee
Taking the lapse $N = |V|$, which is known to simplify the form of the constraint \cite{Ashtekar:2009vc}, and choosing a factor-ordering motivated by previous studies of the Bianchi models in standard LQC \cite{Ashtekar:2009vc, MartinBenito:2009qu}, the various terms have a relatively simple form.  For example,
\begin{align}
\wh{\mC}_{(12)} = & \: \f{\sqrt{|V|}}{64 \pi G \la_m^2} \Bigg[ (\mN_1^+ - \mN_1^-) |V| (\mN_2^+ - \mN_2^-) \nn \\ & + (\mN_2^+ - \mN_2^-) |V| (\mN_1^+ - \mN_1^-) \Bigg] \sqrt{|V|},
\end{align}
and
\be
\wh{\mC}_{j}^n = -\f{|\vt_k \vt_l|^3}{16 \pi G \la_m} \cdot \f{1}{\sqrt{|\vt_j|}} \cdot (\mN_j^+ - \mN_j^-) \cdot \f{1}{\sqrt{|\vt_j|}},
\ee
where it is understood that there is no sum over $j$, and that $k$ and $l$ are both different from $j$ and each other in the definition of $\wh{\mC}_j^n$.  The shift operators $\mN_j^\pm$ appearing here are defined as
\be \label{n-shift}
\mN^\pm_j = \f{1}{2} \Big( \ve_j \mathfrak{N}^\pm_j + \mathfrak{N}^\pm_j \ve_j \Big),
\quad {\rm no~sum~over~} j.
\ee
This factor-ordering, first suggested in \cite{MartinBenito:2009qu}, is convenient since it annihilates any eigenket $|\vt_1, \vt_2, \vt_3\ket$ whose volume $V$ would change sign when acted upon by $\mN^\pm_j$.  Thus, the octants of positive and negative $\vt_j$ are not mixed under the action of $\mN^\pm_j$.  This property will simplify the analysis of the action of the Hamiltonian constraint operator below.

Then, defining the operator
\be \label{theta}
\Theta = -2 \Big[ \wh{\mC}_{(12)} + \wh{\mC}_{(23)} + \wh{\mC}_{(13)} + \lo n^j \wh{\mC}^n_j \Big],
\ee
and requiring that the Hamiltonian constraint operator \eqref{qham} annihilate states $\Psi(\vt_1, \vt_2, \vt_3, \phi)$ in the physical Hilbert space gives
\be
-\hbar^2 \partial_\phi^2 \Psi = \Theta \Psi.
\ee
Since $\Theta$ acts only on the gravitational sector, the scalar field can be used as a relational clock, and $\Theta$ can be treated as a true Hamiltonian (assuming that it is self-adjoint).  The positive frequency solutions, namely the solutions to
\be
-i \hbar \partial_\phi \Psi = \sqrt\Theta \Psi,
\ee
constitute the physical Hilbert space.

At this point, it can be checked that all singular states (i.e., states that correspond to the classical big-bang or big-crunch singularity of $V=0$) are annihilated by $\Theta$.  Denoting singular states as $|\vt_1, \vt_2, \vt_3 \ket_{\rm sing}$,
\be \label{sing}
\Theta |\vt_1, \vt_2, \vt_3 \ket_{\rm sing} = 0,
\ee
and therefore all singular states are stationary with respect to the relational time $\phi$.

An important property of $\Theta$ is that singular states decouple from non-singular states under its action.  First, singular states remain singular, as seen in \eqref{sing}.   Second, any non-singular state that would be shifted to a singular state with $V=0$ is annihilated by a prefactor of $V$ to some power (recall that inverse power operators of $\vt_j$ are defined such that they annihilate states where $\vt_j$ vanishes).  Thus, if one constructs an initial state $\Psi_{ns}(\phi_o)$ that only has support on non-singular states, $\Psi_{ns}(\phi)$ will continue to be non-singular for all $\phi$.

The next step is to study in further detail the action of $\Theta$ on non-singular states.  This task is simplified by the fact that, given the definition of $\mN_j^\pm$ in \eqref{n-shift}, it follows that the octants of positive and negative $\vt_j$ are not mixed under the action of $\Theta$.  Therefore, it is possible to study the action of $\Theta$ on one octant at a time.  A further simplification arises from the parity properties of $\Theta$.  Recall the parity transformations \eqref{parity} (which act on the $\vt_j$ variables as, e.g., $\Pi_1 |\vt_1, \vt_2, \vt_3\ket = |-\vt_1, \vt_2, \vt_3\ket$) under whose action the wave function is required to remain invariant.  It is easy to check that
\be
\Pi_j \mN_k^\pm \Pi_j =
\begin{cases}
\mN_k^\pm & {\rm if~} j \neq k, \\
- \mN_k^\mp & {\rm if~} j = k,
\end{cases}
\quad {\rm no~sum~over~}j,
\ee
from which it immediately follows that
\be
\Pi_j \Theta \Pi_j = \Theta, \quad {\rm no~sum~over~}j.
\ee
Therefore, due to the invariance of $\Theta$ under parity transformations, it is convenient to first determine the action of $\Theta$ on the positive octant where $\vt_j > 0$.  Then, it is easy to determine the action of $\Theta$ in the other octants by exploiting the fact that it is invariant under parity transformations.

When acting on the positive octant, the action simplifies considerably and is given below.  The action of the $\mC_{(jk)}$ terms combine in a relatively simple form, and so the action of $\Theta$ can be written in the following form,
\begin{align}
\Theta \Psi = & \, \f{-\sqrt{V}} {32 \pi G \la_m^2} \Bigg[ V^+ \sqrt{V^{++}}\Psi^{++}
+ \theta_{V^{--}} V^- \sqrt{V^{--}} \Psi^{--} \nn \\ & \!
- \! \sqrt{V} \Big( V^+ \Psi^{-+} \! + \theta_{V^-} \! V^- \Psi^{+-} \Big) \! \Bigg] %\nn \\ &
\! + \lo n^j \Theta_j^n \Psi,
\end{align}
where $\theta_x$ is the Heaviside function which vanishes for $x \le 0$ and is 1 elsewhere, and
\be
V^\pm = V \pm 2 \pi G \hbar \la_m, \qquad V^{\pm\pm} = V \pm 4 \pi G \hbar \la_m.
\ee
The shifted wave functions are defined as
\begin{align}
\Psi^{++}(\vt_1, \vt_2, \vt_3) = \sum_{j \neq k} \mathfrak{N}_j^+ \mathfrak{N}_k^+ \Psi(\vt_1, \vt_2, \vt_3), \\
\Psi^{--}(\vt_1, \vt_2, \vt_3) = \sum_{j \neq k} \mathfrak{N}_j^- \mathfrak{N}_k^- \Psi(\vt_1, \vt_2, \vt_3), \\
\Psi^{+-}(\vt_1, \vt_2, \vt_3) = \sum_{j \neq k} \mathfrak{N}_j^+ \mathfrak{N}_k^- \Psi(\vt_1, \vt_2, \vt_3), \\
\Psi^{-+}(\vt_1, \vt_2, \vt_3) = \sum_{j \neq k} \mathfrak{N}_j^- \mathfrak{N}_k^+ \Psi(\vt_1, \vt_2, \vt_3).
\end{align}
Note that since $\mathfrak{N}_j^\pm$ and $\mathfrak{N}_k^\pm$ don't commute for $j \neq k$, each $\Psi^{\pm\pm}$ contains six terms, and each of them has the form, e.g.,
\be
\mathfrak{N}_1^+ \mathfrak{N}_2^+ \Psi(\vt_1, \vt_2, \vt_3) =
\Psi\Big(\vt_1 \cdot \tf{V^{--}}{V^-}, \vt_2 \cdot \tf{V^-}{V}, \vt_3\Big),
\ee
or
\be
\mathfrak{N}_1^+ \mathfrak{N}_2^- \Psi(\vt_1, \vt_2, \vt_3) =
\Psi\Big(\vt_1 \cdot \tf{V}{V^+}, \vt_2 \cdot \tf{V^+}{V}, \vt_3\Big).
\ee
Importantly, $\Psi^{++}, \Psi^{+-}, \Psi^{-+}$ and $\Psi^{--}$ are all eigenvectors of $\h V$ with eigenvalues $V^{++}, V, V$ and $V^{--}$ respectively.

Finally, the operators $\Theta^n_j$ act as, for example,
\begin{align} \label{theta-n1}
\Theta^n_1 \Psi(\vt_1, \vt_2, \vt_3) = & \,
\f{V^{7/2}}{8 \pi G \la_m \vt_1^4}
\Bigg[ \f{1}{\sqrt{V^+}} \Psi\Big(\vt_1 \cdot \tf{V^+}{V}, \vt_2, \vt_3\Big)  \nn \\ & \:
- \f{\theta_{V^-}}{\sqrt{V^-}} \Psi\Big(\vt_1 \cdot \tf{V^-}{V}, \vt_2, \vt_3\Big) \Bigg].
\end{align}
The actions of $\Theta_2^n$ and $\Theta_3^n$ are given by permutations of \eqref{theta-n1}.  (It is understood that the numerical prefactor $\theta_{V^-} / \sqrt{V^-}$ is zero for $V^- = 0.$)

As already mentioned, the action of $\Theta$ in the different octants are related in a trivial fashion due to the invariance of $\Theta$ under parity transformations.

Note that, as pointed out in \cite{Ashtekar:2009vc}, the explicit form of the action of the Hamiltonian constraint operator can be simplified by using $V$ as a quantum number (which is shifted by a constant factor) rather than one of the $\vt_j$ (which are rescaled by a prefactor that depends on $V$), which should be kept in mind for more detailed analytical or numerical investigations of the quantum dynamics.

This completes the explicit definition of the Hamiltonian constraint operator in self-dual LQC for diagonal type A Bianchi models.  One of its main properties, already mentioned above, is that the singularity is resolved in the sense that non-singular and singular states decouple under its action.

Another interesting property of the Hamiltonian constraint operator is that, when the spatial curvature is non-vanishing, $\Theta$ has a different form in self-dual LQC than in standard LQC.  However, in the case of the Bianchi type I space-time that has zero spatial curvature, the Hamiltonian constraints of self-dual and standard LQC are in fact identical (with the replacement $\gamma \sqrt\Delta \to \la_m$ from \cite{Ashtekar:2009vc}).  Thus, in the absence of spatial curvature, standard and self-dual LQC give the same result, and therefore all of the results obtained in \cite{Ashtekar:2009vc} for standard LQC of Bianchi type I space-times also hold in self-dual LQC.  In particular, there exists a projection map from the self-dual LQC for the Bianchi I space-time to the self-dual LQC for the isotropic flat FLRW space-time which gives the correct Hamiltonian constraint operator, showing that in this restricted setting the symmetry reduction from the Bianchi type I space-time to the flat FLRW space-time (i.e., the imposition of isotropy) commutes with the loop quantization, whether one uses self-dual variables or the Ashtekar-Barbero variables.

\subsection{Relation to Standard LQC}
\label{ss.comp}

Now that the kinematical Hilbert space and the Hamiltonian constraint operators for anisotropic self-dual LQC have been constructed, it is possible to compare the resulting theory with standard anisotropic LQC.  As shall be shown here, while the kinematical Hilbert spaces are isomorphic, the physical Hilbert spaces are different when the spatial curvature is non-zero.  This is the same result that was found for the self-dual LQC of the FLRW space-times \cite{Wilson-Ewing:2015lia}.

In order to compare these two theories, recall that in standard LQC the Ashtekar-Barbero connection and the densitized triads are parametrized as
\be
\t A_a^j = \f{\t c_j}{\lo} \, \ow_a^j, \quad \t E^a_j = \t p_j \, \f{\sqrt{\oq}}{\lo^2} \, \oe^a_j, \quad {\rm no~sum~over~} j,
\ee
where I have placed tildes on the variables of standard LQC in order to differentiate them from the variables of self-dual LQC.  Since $\t A_a^j = \Gamma_a^j + \ga K_a^j$, the relation between these variables is given by
\be \label{comp-c}
\t c_j = -i \ga c_j + (1+i\ga) \lo \Gamma_j, \qquad \t p_j = p_j.
\ee

A convenient basis for the standard LQC kinematical Hilbert space of the Bianchi models is given by eigenstates of the operators corresponding $\t p_j$, whose inner product is given by
\be
\bra \t p_1, \t p_2, \t p_3 | \t k_1, \t k_2, \t k_3 \ket = \de_{\t p_1, \t k_1} \de_{\t p_2, \t k_2} \de_{\t p_3, \t k_3},
\ee
and the other family of basic operators on the kinematical Hilbert space are those corresponding to complex exponentials of $c_j$ which act as shift operators, for example
\be
e^{i \mu c_1} |\t p_1, \t p_2, \t p_3\ket = |\t p_1 + 8 \pi G \hbar \ga, \t p_2, \t p_3\ket.
\ee

Then, the relation \eqref{comp-c} and the Baker-Campbell-Hausdorff equation (to leading order) together suggest relating the operators corresponding to holonomies in standard LQC and generalized holonomies%
\footnote{This relation between standard holonomies of the Ashtekar-Barbero connection and generalized holonomies (with imaginary $\alpha$) of the self-dual connection provides a further motivation to introduce generalized holonomies when working with self-dual variables.}
(with complex-valued $\alpha$) in self-dual LQC via
\be
e^{i \mu \t c_j} = e^{(i - \ga) s \lo / 16 \pi G \hbar \ga} e^{\mu \ga c_j} e^{-(i - \ga) s \lo / 16 \pi G \hbar \ga}.
\ee
If one then also requires that the basis states of standard LQC and self-dual LQC be related by
\be
|p_1, p_2, p_3\ket = e^{(i-\ga) s \lo / 16 \pi G \hbar \ga} |\t p_1, \t p_2, \t p_3 \ket,
\ee
the result is exactly the kinematical Hilbert space defined in Sec.~\ref{ss.kin}, with the same basic operators.  This map provides a simple way to translate between the kinematical Hilbert spaces of standard and self-dual LQC, which are clearly isomorphic.

On the other hand, the Hamiltonian constraint operators in standard and self-dual LQC are different.  This can be seen by taking the Hamiltonian constraint operator of standard LQC for, e.g., the Bianchi IX space-time given in \cite{WilsonEwing:2010rh}, and there replacing the standard LQC operators $\t V = V, \t \vt_j = \vt_j$ and
\be \label{comp-N}
\t \mN_j^\pm = e^{(i - \ga) s \lo / 16 \pi G \hbar \ga} \mN_j^\pm e^{-(i - \ga) s \lo / 16 \pi G \hbar \ga},
\ee
by their counterparts in self-dual LQC.  The resulting Hamiltonian constraint operator is very similar to the self-dual LQC Hamiltonian constraint operator \eqref{theta}, although with two important differences (beyond the factor-ordering ambiguities which in any case arise in both standard and self-dual LQC).  First, the shift operators in the self-dual $\mC_H$ do not preserve the norm of the states they act upon.  This is different from the Hamiltonian constraint operator coming from standard LQC, where the $\mN_j^\pm$ operators are now sandwiched between exponential operators, as seen in \eqref{comp-N}, which ensure that the norm is preserved in that case.  Second, there appear additional terms in the standard LQC Hamiltonian constraint operator coming from the $(1+\ga^2) E^a_j E^b_k \ep^{jk}{}_l \Omega_{ab}{}^l / \sqrt q$ term (with $\Omega_{ab}{}^l = 2 \partial_{[a} \Gamma_{b]}^l + \ep_{jk}{}^l \Gamma_a^j \Gamma_b^k$ corresponding to the spatial curvature) that arises when the Hamiltonian constraint is expressed in terms of the Ashtekar-Barbero variables.  Due to these two differences, the Hamiltonian constraints of standard and self-dual LQC are not equivalent, and the two theories have different physical Hilbert spaces.  This is exactly analogous to what happens for FLRW space-times \cite{Wilson-Ewing:2015lia}.

The main difference between the Hamiltonian constraint operators of the two theories for anisotropic space-times concerns the definition of the non-local connection operator, and whether this basic operator should correspond to the self-dual connection, or the Ashtekar-Barbero connection: this is an ambiguity in the definition of the field strength operator.  Another ambiguity of the same type arises for the loop quantization of the closed FLRW space-time, where it is possible to directly define a non-local field strength operator, or instead first define a non-local connection operator and use this operator to define the field strength operator.  An interesting result is that, while there do exist some quantitative differences between the dynamics resulting from the different definitions of the field strength operator, the qualitative behaviour is not affected by this ambiguity \cite{Singh:2013ava, Corichi:2011pg, Corichi:2013usa}.  Based on this result, it appears possible that this other type of field strength quantization ambiguity ---namely, whether the non-local connection operator corresponds to the self-dual connection or the Ashtekar-Barbero connection--- will not significantly affect the qualitative predictions of anisotropic LQC.

This is reasonable since, despite the differences between their Hamiltonian constraint operators, there are important similarities between standard LQC and self-dual LQC.  Most obviously, the Hamiltonian constraint equation has the form of a difference equation, and the big-bang and big-crunch singularities of the classical theory are resolved in the sense that the zero-volume states decouple from the non-singular states under the action of the Hamiltonian constraint operator.

Some further evidence in this direction is offered by the following.  In the limit of vanishing spatial curvature, the resulting space-time is the Bianchi type I model.  For this model, all of the $n^j=0$ and hence $s=0$, and as pointed out above in Sec.~\ref{ss.ham}, the self-dual and standard LQC Hamiltonian constraints are in this case equal.  Therefore, the predictions of self-dual and standard LQC will agree in regimes where the spatial curvature can be neglected.  Nonetheless, numerical simulations are likely necessary in order to quantify the differences between the quantum dynamics of standard and self-dual LQC, especially in the Planck regime.

The main open question at this point is to understand the full quantum dynamics, first for semi-classical states, and then for more widely spread states (as has been done in standard LQC for isotropic space-times in \cite{Ashtekar:2006wn, Ashtekar:2007em, MenaMarugan:2011me} and \cite{Diener:2013uka, Diener:2014mia, Diener:2014hba}, respectively).  However, this is a very difficult problem for the Bianchi space-times due to the complexity of the difference equation coming from the Hamiltonian constraint operator, and particularly of the form that the shifts in the $\vt_j$ variables takes.  Indeed, there do not yet exist any studies of the quantum dynamics of semi-classical states in standard LQC for the Bianchi type I space-time (for the improved dynamics prescription given in \cite{Ashtekar:2009vc}), the simplest of the Bianchi models.  Nonetheless, despite the difficulty of studying the full quantum dynamics of even semi-classical states, important insights can be obtained by studying the effective equations.

\section{The Effective Theory}
\label{s.eff}

The effective theory is obtained by treating the Hamiltonian constraint operator \eqref{qham} as a classical constraint on the original phase space, which is then called the effective Hamiltonian constraint \cite{Ashtekar:2006wn, Taveras:2008ke}, and the effective equations are simply given by the Poisson brackets of the observable $\mO$ of interest with the effective Hamiltonian constraint, $d\mO / dT = \{\mO, \mC_H^{\rm eff}\}$.

The effective equations have already been found to be very useful in standard LQC where they are in good agreement with the full quantum dynamics of a large class of semi-classical states.  Indeed, in the isotropic space-times in standard LQC where the full quantum dynamics of semi-classical states have been studied analytically and numerically, the effective dynamics provide an excellent approximation to the evolution (with respect to a relational clock) of expectation values for semi-classical states that (i) are sharply peaked in both conjugate variables and (ii) where the total volume of the space-time remains much larger than the Planck volume $\lp^3$ \cite{Ashtekar:2006wn, Ashtekar:2006es, Bentivegna:2008bg, Pawlowski:2011zf, Pawlowski:2014fba, Diener:2014mia}.  While the reliability of the effective equations may appear surprising at first, it can be understood to arise due to the fact that the variables of interest in quantum comsology (the total volume $V$, the total momentum of the scalar field $\pi_\phi$, etc.) are global observables and correspond to heavy degrees of freedom so long as $V \gg \lp^3$.  Therefore, quantum fluctuations do not become important (assuming they are initially small), and the effective equations can be trusted even at the bounce point where quantum gravity effects are strongest \cite{Rovelli:2013zaa}.

Thus, the effective equations are also expected to provide a good approximation to the quantum dynamics of sharply peaked states in anisotropic self-dual LQC so long as all $\vt_j$ remain much larger than $\lp$.  Following the procedure outlined above, the effective Hamiltonian constraint ---for all diagonal type A Bianchi models in self-dual LQC--- is
\begin{align}
\mC_H^{\rm eff} \! =& \, \f{N V}{8 \pi G \la_m^2}
\Bigg[ \sinh \f{b_1}{\vt_2 \vt_3} \sinh \f{b_2}{\vt_1 \vt_3} \nn \\ &
- \f{\lo n^1 \la_m V}{\vt_1^4} \sinh \f{b_1}{\vt_2 \vt_3}
+ \sinh \f{b_1}{\vt_2 \vt_3} \sinh \f{b_3}{\vt_1 \vt_2} \nn \\ & 
- \f{\lo n^2 \la_m V}{\vt_2^4} \sinh \f{b_2}{\vt_1 \vt_3}
+ \sinh \f{b_2}{\vt_1 \vt_3} \sinh \f{b_3}{\vt_1 \vt_2} \nn \\ & 
- \f{\lo n^3 \la_m V}{\vt_3^4} \sinh \f{b_3}{\vt_1 \vt_2} \Bigg]
\label{eff-ham}
+ \f{N p_\phi^2}{2 V} \approx 0.
\end{align}
Here the lapse $N$ has been left free and $\mC_H^{\rm eff}$ is given for $\vt_j > 0$ (in the effective theory, just as in the classical theory, it is possible to work in just one octant).

Now the effective equations can be calculated in a straightforward fashion.  For example,
\begin{align}
\f{\d \vt_1}{\vt_1} = \, & -\f{i N}{2 \la_m^2} \bigg[ 
\sinh \f{b_2}{\vt_1 \vt_3} + \sinh \f{b_3}{\vt_1 \vt_2}  \bigg] \cosh \f{b_1}{\vt_2 \vt_3} \nn \\ &
- \f{i \lo n^1 N V}{2 \la_m \vt_1^4} \cosh \f{b_1}{\vt_2 \vt_3},
\end{align}
with the effective equations for $\d\vt_2$ and $\d\vt_3$ being given by the appropriate permutations.

These can be related to the directional Hubble rates, which are given by, e.g.,
\be
H_1 = \f{\d a_1}{a_1} = \f{\d\vt_2}{\vt_2} + \f{\d\vt_3}{\vt_3} - \f{\d\vt_1}{\vt_1},
\ee
and the mean Hubble rate is given by
\be
H = \f{\d a}{a} = \f{1}{3} \left( H_1 + H_2 + H_3 \right),
\ee
where $a = (a_1 a_2 a_3)^{1/3}$.

In some cases in standard LQC, it has been possible to bound the directional Hubble rates in the effective theory (and hence bound the expansion and the shear also) \cite{Corichi:2009pp, Gupt:2011jh}.  These bounds then provide evidence that all strong singularities are resolved in these space-times in standard LQC \cite{Singh:2009mz}.  What can be said regarding the effective theory of self-dual LQC?

As already mentioned in Sec.~\ref{ss.ham}, for the spatially flat Bianchi type I space-time (where the spin-connection $\Gamma_a^j$ vanishes), standard LQC and self-dual LQC give the same resulting quantum theory, and hence also the same effective dynamics.  Therefore, the results already found by studying the effective equations of the Bianchi I model in standard LQC also hold for self-dual LQC.  In particular, the expansion and shear both have an upper bound around the Planck scale \cite{Corichi:2009pp}, and Kasner transitions of the type found in \cite{Gupt:2012vi} will occur at the bounce.

Concerning the dynamics of Bianchi space-times with non-vanishing spatial curvature in the effective theory coming from self-dual LQC, it is likely that numerical simulations like the ones recently performed in standard LQC for the Bianchi II and Bianchi IX space-times \cite{Corichi:2012hy, Corichi:2015ala} will be necessary in order to understand the self-dual LQC dynamics of these space-times in all regimes.  Nonetheless, it is known from general relativity that as a big-bang or big-crunch singularity is approached in the Bianchi models with a massless scalar field (the case considered here) in a fashion where the $\d a_j$ all have the same sign (i.e., in general relativity this would lead to what is called an isotropic or `point-like' singularity in the terminology of \cite{Thorne:1967zz}), the space-times become asymptotically velocity-term dominated (AVTD) in which case the spatial curvature is negligible \cite{Berger:2002st}.  When the spatial curvature becomes negligible, then the Hamiltonian constraint is essentially that of the Bianchi I space-time, and all of the results obtained for Bianchi I, described in the paragraph above, hold for AVTD Bianchi space-times also.

On the other hand, if some of the $\d\vt_j$ have different signs as the singularity is approached (which would give, e.g., a cigar-like or a pancake-like singularity in general relativity), then a Bianchi space-time with non-vanishing curvature will not be of the AVTD type and one cannot rely on results coming from the study of the Bianchi I space-time.  While considerably more work is needed in order to study the dynamics of non-AVTD Bianchi space-times, this case is particularly interesting as it is here that the chaotic mixmaster behaviour of the Bianchi IX space-time arises in general relativity.  In particular, it would be interesting to check in detail whether the chaotic behaviour found classically persists in the effective theory.  While there are indications that the chaotic behaviour may disappear when quantum gravity effects are included \cite{Bojowald:2003xe, Bojowald:2004ra}, this question lies outside of the scope of this paper and is left for future work.

\section{Discussion}
\label{s.disc}

A loop quantization of the diagonal type A Bianchi space-times in terms of the self-dual connection variables has been presented in this paper.  The reality conditions were imposed through the choice of the inner product in the kinematical Hilbert space, and then the `improved dynamics' Hamiltonian constraint operator was constructed using a non-local connection operator (as is also done in standard LQC for anisotropic space-times with non-vanishing spatial curvature).  A more detailed study of the dynamics of semi-classical states is left for future work.  Note that while using self-dual variables simplifies the form of the Hamiltonian constraint operator since the second term (that appears when using Ashtekar-Barbero variables) with the prefactor $(1+\ga^2)$ vanishes for $\ga=i$, this happens at the cost of an inner product that has a more complicated form.  At this point, it is not clear whether calculations in the quantum theory are easier with a simple inner product and a more complicated Hamiltonian constraint operator, or vice versa.

The key step in this construction was imposing the reality conditions.  This was done in two parts: first the reality conditions were expressed in terms of the fundamental operators of the theory, with the relation \eqref{real} simplifying the task, and second, the results of the study of the closed FLRW space-time in terms of self-dual variables motivated an ansatz for the form of the inner product which was then shown to correctly implement the reality conditions.  The fact that this could be done in a relatively simple fashion for all diagonal type A Bianchi models raises the hope that this may also be possible in more general settings.

Another important point, already noticed in the study of the self-dual LQC of FLRW space-times, is the necessity of introducing a family of generalized holonomies parametrized by $\alpha \in \mathbb{C}$.  It turns out that it is only when $\alpha$ is purely imaginary that the generalized holonomies are well defined in self-dual LQC --- standard holonomies of $A_a^j$ (for which $\alpha = 1$) are not well defined in the kinematical Hilbert space.  Essentially, the generalized holonomies of interest (i.e., the ones that become fundamental operators in self-dual LQC) correspond to objects of the type $h \sim \mP \exp \int i A_a$.

The reason that operators of this type are the ones that are well defined in the quantum theory can be understood in the following manner.  Generalized holonomies of the complex-valued Ashtekar connection cannot be self-adjoint, no matter the choice of $\alpha$.  Then, speaking loosely, the question becomes whether one prefers the extrinsic curvature part to be self-adjoint, or the spin-connection part to be self-adjoint.  Recalling that the holonomy of the real-valued Ashtekar-Barbero connection is a self-adjoint operator in standard LQC, and since it is the extrinsic curvature which is canonically conjugate to the densitized triad, this suggests choosing the extrinsic curvature part of the generalized holonomy to be self-adjoint, and this corresponds to a purely imaginary $\alpha$.  From a more technical perspective, this issue is settled by the fact that for the shift operator to be well defined in the kinematical Hilbert space of self-dual LQC, the shift must be real-valued, and this requirement also constrains $\alpha$ to be purely imaginary.

An important question now is what can be learnt from this process and applied to full LQG.  First, these results offer the hope that it may be possible to properly implement the reality conditions, and in particular some of the techniques developed here may prove to be useful more generally in imposing the reality conditions.  Second, the necessity of using generalized holonomies of the form $h \sim \mP \exp \int i A_a$ in self-dual LQC suggests that this type of generalized holonomies may play an important role in self-dual LQG as well.  Note however that an important property of standard holonomies is that under gauge transformations they behave in a very simple fashion.  One difficulty in using generalized holonomies in full quantum gravity (where there are no natural gauge-fixings available as in quantum cosmology minisuperspaces) is that they behave in a significantly more complicated fashion under gauge transformations.  Clearly, this is a difficulty that must be addressed in order to define a version of self-dual LQG based on generalized holonomies.  The important problem of whether it is possible to construct a well defined version of self-dual LQG based on generalized holonomies is left for future work.

Note that another important open problem that must be addressed in order to properly define self-dual LQG is to have a well defined measure on the space of generalized connections.  The measure problem is avoided in self-dual LQC due to the symmetry reduction that is performed before quantization, and therefore there are no direct lessons to be learnt from self-dual LQC regarding this last open problem.

Nonetheless, there are some hints that a careful implementation of the reality conditions may be important for the solution of the measure problem in self-dual LQG.  A na\"ive choice for the inner product in self-dual LQC motivated by the inner product of standard LQC, before any consideration of the reality conditions, might be of the form
\be \label{naiveIP}
\bra \psi_1 | \psi_2 \ket \! = \!\! \lim_{L \to \infty} \f{1}{2L} \int_{-L}^L \int_{-L}^L \!\! dx \, dy \, \: \overline{\psi_1 \! (c)} \: \psi_2 \! (c),
\ee
where $x = {\rm Re}(c)$ and $y = {\rm Im}(c)$.   (Here for the sake of simplicity I have considered the isotropic case where the gravitational configuration space is one-dimensional; however this na\"ive inner product can directly be generalized for the diagonal type A Bianchi models.)  This inner product, however, is pathological as it gives a divergent norm for the eigenkets $|p\ket \sim e^{p c}$.  Thus, in a (na\"ive) sense there is a `measure problem' in self-dual LQC also.  However, once the reality conditions are taken into account, then the resulting inner product is sufficiently different from \eqref{naiveIP} that it gives a finite norm for the eigenkets $|p\ket$ and has the main qualitative properties expected of an LQC inner product (see \cite{Wilson-Ewing:2015lia} for details).  It is possible that a similar result could be found in LQG: the problems of the reality conditions and of the non-compact measure of the generalized connection may not be disconnected.  Rather, this argument suggests that it could be productive to work on both problems simultaneously and that the solution of one problem may perhaps simultaneously provide a solution to the other.

\medskip

\acknowledgments

I would like to thank Joseph Ben Geloun and Casey Tomlin for helpful discussions.
This work is supported by a grant from the John Templeton Foundation.

%\bibliographystyle{bib-style}
%\bibliography{bibliography}

%\end{document}

\raggedright

\end{document}